\documentstyle[preprint,aps,eqsecnum]{revtex}

\begin{document}
\draft
\preprint{IFT/17/94}
\tighten

\title   {     Quadrupole collective states in a large single-$j$ shell
         }
\author  {     K. Burzy\'nski$^1$ and J. Dobaczewski$^{1,2}$
         }
\address {             $^1$Institute of Theoretical Physics,
                                Warsaw University,
                         Ho\.za 69, PL-00-681 Warsaw, Poland\\
                           $^2$University of Pittsburgh,
                            Pittsburgh, PA 15260, USA
         }

\maketitle

\begin{abstract}
We discuss the ability of the generator coordinate method (GCM) to
select collective states in microscopic calculations.
The model studied is a single-$j$ shell with hamiltonian containing
the quadrupole-quadrupole interaction.
Quadrupole collective excitations are constructed
by means of the quadrupole single-particle operator.
Lowest collective bands for $j$=31/2 and particle numbers
$N$=4,6,8,10,12, and $14$ are found. For lower values of $j$,
exact solutions are obtained and compared with
the GCM results.
\end{abstract}


\narrowtext

\section{Introduction}
\label{sec1}

A microscopic description of nuclear collective modes
is still a very important and interesting problem studied in
the theory of nuclear structure.
In a frame of the shell model such a description
is difficult to obtain
because of a very large size of the basis required to include all
necessary configurations. For example, only very recently
excellent shell-model results for rotational states in $^{48}$Cr
have been obtained \cite{CZP94}.
Although many phenomenological approaches can, and have been
constructed,
their microscopic derivation from the shell model is often not
entirely accomplished.
It is therefore important to analyze and study approaches
which are able to provide
a simultaneous description of collective and single-particle phenomena.

The generator coordinate method (GCM) \cite{HW53,OY66,Won75,RS80}
constitutes an effective method to deal
with collective degrees of freedom
while starting from a pure many-fermion description.
Moreover, it may also yield an exact quantum mechanical
formalism, and only the choice of generating functions and a generator
coordinate decides whether we can find exact solutions, collective as
well as noncollective, to the actual problem.  The configuration
mixing feature which is by construction built in into the GCM
provides us with the ideal tool to study microscopic
foundations of collective models.

In the present study, we apply the GCM to a
microscopic model of
a single-$j$ shell filled with even number of identical nucleons interacting
via pairing-plus-quadrupole hamiltonian \cite{Bel59}.
Although the model is very simple, it contains two main ingredients
of the ``real'' nuclear structure theories, i.e., the quadrupole deformation
and the pairing correlations, and at the same time for
low values of $j$ it is relatively easy to solve.
On the other hand, the GCM is well suited to study this model
for large values of $j$ where the exact solutions are inaccessible.

In the past the single-$j$ shell model has been
studied by many authors.
Mulhall and Sips \cite{MS64} found that even for 4 particles occupying
single-$j$ shell with strong quadrupole-quadrupole force the collective
effects
are important and rotational structures are present in the spectra.
Deformability, i.e., a competition between the quadrupole force and
the pairing,
was studied in a series of papers by Baranger and
Kumar \cite{BK65,BK68}.
Arima \cite{Ari68} discussed excited rotational level structures ($K$=2
bands) appearing in the single-$j$ shell.
Friedman and Kelson \cite{FK70} analyzed collective
spectra as depending on particle number and $j$.
The present paper aims at using the GCM as a filter which would
select collective structures among the complete shell-model spectrum.
Such a study may give us more confidence in using the GCM in more realistic
cases where the quality of approximations involved is difficult to analyze.

The paper is organized in the following way:
In Sec.{\ }\ref{sec1a} we fix notation by briefly reviewing the
pairing-plus-quadrupole model in a single-$j$ shell
and then in Sec.{\ }\ref{sec4a} we present the available
exact solutions.
The key point of the GCM is the construction of the
generating states, which we present in Sec.{\ }\ref{sec1b}
by invoking the
single particle coherent excitation model (SCEM) of Dobaczewski
and Rohozi\'nski \cite{Dob90,DR91}. In this model the
quadrupole excitations are
build using the single-particle quadrupole operator.
In the present study we restore the broken
particle-number and angular-momentum symmetries exactly.
This is done in a consistent GCM framework by using the gauge angle
and the Euler angles as generator coordinates.
The particle-number projection \cite{Bay60,DMP64} is based on
the Fomenko method \cite{Fom70,Sor72} in which integrals are replaced
by finite sums, Sec.{\ }\ref{sec2},
and a similar method is also used to perform
the angular momentum projection \cite{RW54,PY57},
Sec.{\ }\ref{sec1c}.
A description of the GCM calculations in the intrinsic frame of
reference is presented
in Sec.{\ }\ref{sec3}.
In Sec.{\ }\ref{sec4b} we analyze the GCM results for
$j$=15/2 and $N$=8 particles and compare them with the exact calculations.
In Sec.{\ }\ref{sec4c} we present the GCM spectra for $j$=31/2
and $N$=4, and for
particle numbers between 6 and 14,
where the exact solutions are not available.

\section{Single-\protect\mbox{$j$} shell}
\label{sec1a}

We consider the $(2j$+1)-fold degenerate single shell of angular
momentum $j$ filled with even number $N$ of identical particles,
which without the interaction is assumed to be at zero
energy.  The hamiltonian is composed of the pairing-plus-quadrupole
interaction,
\begin{equation} \label{s1}
    \hat H = - G \hat P^+ \hat P  -  \chi \hat Q \cdot \hat Q \;,
\end{equation}
where $\hat P^+$ is the pair transfer operator and $\hat Q$ is
the quadrupole moment operator,
\begin{mathletters}\begin{eqnarray}
   \hat P^+ &=& \sum_{mm'}(jmjm'|00) a_m^+ a_{m'}^+\;,
          \label{a2b} \\
   \hat Q_\mu^+ &=& \sum_{mm'}(jmjm'|2\mu) a_m^+ \tilde a_{m'}\;,
          \label{a2a}
\end{eqnarray}\end{mathletters}\noindent%
while $G$ and $\chi$ are pairing and quadrupole coupling
constants, respectively.  Hamiltonian (\ref{s1}) describes basic
collective correlations between nucleons \cite{Bel59,MS64} and
it has been used by many authors
\cite{BK65,BK68,Ari68,FK70,TOA71,VD71}.

In the present paper we aim at solving the pairing-plus-quadrupole
model in a large single-$j$ shell. Dimensions of the
many-fermion space increase rapidly with $j$ and the exact
solutions become inaccessible very fast. For example, in the $M$
and $J$ representations the maximum dimensions for $j$=15/2 are
$526$ and $35$, respectively, while for $j$=31/2 they are as
large as $8\,908\,546$ and $200\,691$.  The generator coordinate method,
which we apply in Sec.{\ }\ref{sec3} to this problem, is able to describe
low lying collective states without using very large matrices.
Here we present solutions
for the values of particle number $N$ and
single-particle angular momentum $j$ for which in the $M$ representation
the dimensions of
the many-particle space are not larger than 5000.

\subsection{Exact diagonalization}
\label{sec4a}

In Fig.{\ }\ref{f11} we show the yrast bands for $N$=8 particles
in the $j$=15/2 shell obtained with several values of the coupling
constants $G$ and $\chi$, see Eq.{\ }\ref{s1}.
Using the parametrizations
$G$=(1$-$$x)\times$1~MeV and $\chi$=$x\times$1~MeV we show
a transition from the pairing (seniority) limit at $x$=0 to a
large-deformation limit at $x$=1. It can be seen that for $x$$<$1 the
spectra are strongly influenced by the fixed-seniority
structures which at $x$=0 appear as degenerate multiplets. Since
in this paper we want to discuss situations corresponding to
collective quadrupole excitations, in the following we present
only the results for the pure quadrupole-quadrupole force, i.e.,
we set $G$=0 and $\chi$=1~MeV.

In Fig.{\ }\ref{f13}(a) we present the yrast bands for $N$=4
particles interacting with the quadrupole-quadrupole interaction
in shells corresponding to $j$ between 15/2 and 31/2.
At low spins, regular rotational-like bands are obtained
with moments of inertia ${\cal J}$=$3\hbar^2\!/E_{2^+}$ varying
between
51~$\hbar^2\!$/MeV for $j$=15/2 and
333~$\hbar^2\!$/MeV for $j$=31/2,
and the corresponding $E_{4^+}/E_{2^+}$ ratios varying between
3.23 and 3.32 ($E_{I^+}$ denotes the excitation energy above
the ground state).

At higher spins the ground-state bands are crossed by bands
built on high-$K$ particle-hole excitations, which can easily
be identified with oblate structures. In the Nilsson single-particle
diagram
presented in Fig.{\ }\ref{f16} for $j$=15/2, we see that
at the oblate side the four particles occupy the
$\Omega$=15/2 and 13/2 orbitals. Therefore the lowest single-particle
excitation corresponds to promoting a particle from $\Omega$=13/2 to
$\Omega$=11/2 which gives the $K$=13/2+11/2=12 excitation and
leads to a band head at $I$=12.
Similarly, for larger values of $j$ one obtains band heads
at larger even values of spin.

For $N$=6 particles, Fig.{\ }\ref{f13}(b), we see the
bands which correspond to 2p-2h oblate configurations. For example,
for $j$=21/2 the p-h band corresponds to exciting the
$\Omega$=17/2 particle to the $\Omega$=15/2 orbital ($K$=16)
and then the second p-h excitation aligns the
$\Omega$=19/2 particle to the $\Omega$=13/2 orbital leading
to $K$=32 and giving the band head at $I$=32.
We see that for small shell-filling factors $N$/(2$j$+1) the
high-$K$ bands appear at very low energies, while they cease
to be yrast for larger shell-filling factors. For example,
for $N$=8 and 10, Figs.{\ }\ref{f13}(c) and (d), no yrast
high-$K$ bands are seen and only yrast-yrare band interaction
is visible at $I$=12 for the lowest $j$=19/2 and $N$=8 bands.
In general, the yrast bands for half-filled shells present regular
collective structures, although they do not have pure rotational
character, for example,
$E_{4^+}/E_{2^+}$=1.92 for the $j$=$N$$-$1/2=15/2 band.
This is due to the $\gamma$-instability of the corresponding
potential energy surface
(PES), see Fig.{\ }\ref{f2}(a).

The complete spectrum of 526 states for $N$=8 particles in
the $j$=15/2 shell is presented in Fig.{\ }\ref{f14}.  The
lines connect states with the largest E2 reduced matrix elements of the
quadrupole operator. Only the stretched $\Delta I$=2 matrix elements
are considered. The lowest $0^+$ state is connected to the $2^+$ state
with which it has the largest E2 matrix element. Then the
first excited $0^+$ state is connected to one of the remaining $2^+$
states, and the procedure is continued for still higher $0^+$ states.
After that all remaining $2^+$ states are considered to be the band heads.
In the next step the $2^+$ states are considered by order of increasing
energy and connected to the $4^+$ states. In this way all even spins
are connected into bands and then the same procedure is
repeated for odd spin values.

Several regular collective bands can be seen in Fig.{\ }\ref{f14}.
These will be used to discuss the ability of the GCM to select
collective structures in the sea of all possible eigenstates,
Sec.{\ }\ref{sec4b}. Here we only discuss a possible interpretation
of lowest states in terms of the collective model.
In Table~\ref{t5} we present the reduced E2 matrix elements
between five lowest $0^+$ and $2^+$ states. It can be seen that
these matrix elements obey a strong selection rule related to the
special symmetry of the half-filled shell. Indeed, the quadrupole operator
(\ref{a2a}) is odd with respect to changing the creation operator
into the annihilation operator and {\it vice versa}. Therefore,
the quadrupole-quadrupole hamiltonian is invariant with respect
to such a change and the spectra of systems containing
$N$ and 2$j$+1$-$$N$ particles are strictly identical. However,
for a half-filled shell $N$ is equal to 2$j$+1$-$$N$ and the above
symmetry allows to attribute a new dychotomic quantum number to every
eigenstate, while the quadrupole operator may only connect states
with these quantum numbers being different.

An approximate interpretation of the lowest states can be done
in terms of the
collective Jean-Wilets model \cite{JW56}, which describes collective states
for the $\gamma$-unstable PES. In this model
the lowest-order quadrupole transition operator changes the
seniority quantum number by one and therefore the parity
of the seniority can be associated with the dychotomic quantum number
discussed above. In this way the $0^+_1$ and $0^+_2$ states can be
identified with seniority zero and three and the three lowest
$2^+$ states with seniority one, two, and four. In this scheme
the states $0^+_3$ and $0^+_4$ can be interpreted as mixtures of the
seniority six states with the seniority zero $\beta$ vibrations,
and the states $2^+_4$ and $2^+_5$ as mixtures of seniority
five and seven with seniority one $\beta$ vibrations. This
interpretation is however not manifested in the energy spectra,
where the characteristic Jean-Wilets multiplets are strongly split.
One can also note that if the identification of the $\beta$
vibrations is correct, the corresponding energy is rather high as
compared to the energy of the motion in
the $\gamma$ direction. This high energy may also suggest that
the collective $\gamma$ vibrations are here directly mixed
to non-collective states and that the $\beta$ vibrations are absent
in the model.

The first excited band can be interpreted as the $\gamma$
vibration strongly coupled to rotations due to the $\gamma$-instability
of the corresponding PES, Fig.{\ }\ref{f2}(a).
At spin $I$=8 it is perturbed by a coupling to the lowest
noncollective excitation promoting the $\Omega$=7/2
particle to the $\Omega$=9/2 state (if one considers
the oblate side of the Nilsson diagram,
Fig.{\ }\ref{f16}). However, one is here unable to distinguish
between the oblate and prolate structures, because the
prolate side suggests an excitation of the $\Omega$=9/2
particle to the $\Omega$=7/2 state, which gives the same value
of spin $I$=8.

For $j$=31/2 the exact solutions are available for $N$=4 only,
Fig.{\ }\ref{f10}. The
collective bands are identified by the same procedure
as described above using
the reduced matrix elements of the quadrupole
operator.
At low spins one can see several regular bands
which can be fairly well interpreted in terms of the $\gamma$ vibrations
coupled to an axial rotor with $K$ being approximately a good quantum number.
In this interpretation, the ground-state band is the
$K$=0 and $n_{\gamma}$=0 band and the first excited band has
$K$=2 and $n_{\gamma}$=1. Then come two $n_{\gamma}$=2 bands
with $K$=4 and $K$=0 followed by two $n_{\gamma}$=3 bands
with $K$=6 and $K$=2. One may even identify three $n_{\gamma}$=4
bands with $K$=8, 4, and 0. Again, in this interpretation there
is no room for the $\beta$ vibrations at low energies \cite{CB94}.
Table~\ref{t6} shows the E2 reduced matrix elements between the lowest
$0^+$ and $2^+$ states and confirms the above interpretation.
A clear decrease of the matrix elements with the difference in the
number of $\gamma$-vibrational quanta is manifest, however,
the strict $|\Delta n_{\gamma}|$$\leq$1 selection rule is not present.

Similarly as for the yrast bands, Fig.{\ }\ref{f13}, one may
easily recognize in the spectrum of  Fig.{\ }\ref{f10} the bands
built on the oblate particle-hole excitations. Together
with the excitation of the $\Omega$=29/2 particle to the
$\Omega$=27/2 orbital, which becomes yrast at $I$=28, one
can also identify the excitations of the same particle to the $\Omega$=25/2,
23/2, 21/2, and 19/2 orbitals.
One may clearly see the effect of decreasing
E2 probability when the band terminates; the band heads are then
coupled stronger to adjacent bands then to the next members of the given band.

\subsection{The SCEM model in a single-$j$ shell}
\label{sec1b}

The single-particle coherent excitation model (SCEM) proposed by
Dobaczewski and Rohozi\'nski \cite{Dob90,DR91} is based on the
assumption that the quadrupole excitations can be constructed by
acting with the quadrupole fermion operator $\hat F^+$ on a
reference state of the form
\begin{equation} \label{a1a}
   \mid\!\text{ref}\,\rangle = \bigl(\hat S^+\bigr)^{N/2}\mid\!0\,\rangle\;,
\end{equation}
i.e.,
\begin{equation}\label{a1b}
  \mid\!\Psi_{\text{SCEM}}\,\rangle
   = \bigl(\hat F^+\bigr)^{N_F}\mid\!\text{ref}\,\rangle\;,
\end{equation}
where $\hat S^+$ creates a monopole collective pair of particles
and $\hat F_\mu^+$ is a single-particle quadrupole excitation
operator.  Since only one operator of a given mutipolarity can
be constructed in the single-$j$ shell, here $\hat
S^+$$\equiv$$\hat P^+$ and $\hat F^+$$\equiv$$\hat Q$.  In the
SCEM, for $N_F$=0 we obtain one state of $I$=0 and for $N_F$=1
one state of $I$=2. If for every value of $N_F$ we orthogonalize
the SCEM states with respect to those obtained for $N_F$$-$1, we
obtain the numbers of states listed in Table~\ref{t1} for
$j$=15/2 and $N$=8.

It can be seen that for low values of $N_F$ the numbers of
states of a given $I$ exactly correspond to those of the
five-dimensional harmonic oscillator \cite{BM75} with $N_F$ phonons, or to
the numbers of $d^+s$ excitations required to create the
interacting-boson-model states \cite{AI75} from the s-boson condensate
$(s^+)^{N_F}$.  However, SCEM being a true fermion model, there
are departures from this rule appearing as soon as $N_F$ reaches
half of the shell's maximum particle number $j$+1/2
(i.e., in Table~\ref{t1} from
$N_F$=4 on), which reflect the effect of the Pauli correlations.
It is obvious that states of the angular momentum $I$ can only
be obtained for $N_F$$\geq$$I/2$ and therefore the number of
excitations $N_F$ plays the role of the cut-off in spin.

The generating function of the GCM is here constructed as a
condensate of the $\hat F^+$ excitations acting on a condensate
of the $\hat S^+$ pairs
\begin{equation} \label{a3}
     \mid\!t_0 t_2\rangle = \exp\bigl(t_2\cdot\hat F^+\bigr)
          \exp\bigl(t_0\hat S^+\bigr) \mid\!0\,\rangle \;.
\end{equation}
The parameters $t_0$ and $t_2$ are a scalar and a quadrupole
numerical tensors, respectively, and are used as the generator
coordinates.  We use the notation of the scalar product defined
as \mbox{$t_2$$\cdot$$\hat F^+$=$\sum_\mu
t_{2\mu}^{\displaystyle\ast} \hat F_\mu^+$}
and assume that
$t_{2\mu}^{\displaystyle\ast}$=$(-1)^{\mu}t_{2,-\mu}$
and $t_0$=$t_0^{\displaystyle\ast}$.

For the GCM state (\ref{a3}) neither the particle number nor the
angular momentum is a good quantum number.  It can, however, be
decomposed into the SCEM states (\ref{a1b}) which, with a proper
angular momentum coupling of the excitation operators
$\hat F^+$, have good angular momentum and particle number.  The
average value of the particle number in the generating function
(\ref{a3}) increases with the $t_0$ parameter. Similarly, the
average number of the excitations $\hat F^+$, and hence the
average value of the angular momentum, increases with the
magnitude of the quadrupole tensor \cite{beta}
\begin{equation}
     \beta^2 = \sum_\mu |t_{2\mu}|^2 \;.
\end{equation}

It is convenient to introduce three matrices
\begin{mathletters}\begin{eqnarray}
 && S_{mm'}^+     = (jmjm'|00) = (-1)^{j-m}\frac{1}{\sqrt{2j+1}}
                            \delta_{m,-m'}\;,           \label{a4a}\\
 &&F_{mm'}^{\mu+} = (jmj\,{-m'}|2\mu) (-1)^{j+m'} ,     \label{a4b}\\
 && F_{mm'}^+     = \sum_\mu t_{2\mu}^{\displaystyle\ast} F_{mm'}^{\mu+}\;,
                                                \label{a4c}
\end{eqnarray}\end{mathletters}\noindent%
where $S$ is antisymmetric and $F$ is hermitian, in terms of
which the basic SCEM building blocks read
\begin{mathletters}\begin{eqnarray}
   \hat S^+ &=& \sum_{mm'}S_{mm'} a_m^+ a_{m'}^+\;,
          \label{a2aa} \\
   \hat F_\mu^+ &=& \sum_{mm'}F_{mm'}^{\mu+} a_m^+ a_{m'}\;,
          \label{a2ab} \\
   t_2\cdot\hat F^+ &=& \sum_{mm'}F_{mm'} a_m^+ a_{m'}\;.
          \label{a2ac}
\end{eqnarray}\end{mathletters}\noindent%
Now, the generating function (\ref{a3}) can be represented as
\begin{equation} \label{a5}
   \mid\!t_0 t_2\rangle = \exp\bigl( t_0 e^{t_2\cdot \hat F^+}
        \hat S^+  e^{-t_2\cdot \hat F^+} \bigr) \mid\!0\,\rangle \;,
\end{equation}
and further, using the nonunitary Bogoliubov transformation
\cite{BB69}, as a Thouless \cite{Tho60} state:
\begin{equation} \label{a6}
  \mid\!t_0 t_2\rangle = \exp\biggl( \case{1}{2} \sum_{mm'} C_{mm'}^+
                  a_m^+ a_{m'}^+ \biggr) \mid\!0\,\rangle \;,
\end{equation}
where
\begin{equation} \label{a8}
   C_{mm'}^+=2 t_0 \sum_{nn'} \bigl( e^{F} \bigr)_{mn} S_{nn'}^+
                            \bigl( e^{F^{\displaystyle\ast}} \bigr)_{n'm'}\;.
\end{equation}
In this way all GCM kernels can be easily calculated using the
standard expressions resulting from the Wick theorem
\cite{Wic50,RS80}.  On the other hand, the product of matrices
in (\ref{a8}) can be rapidly obtained by using the $F$ matrix in
its eigen-reference-frame, which corresponds to transforming the
tensor $t_2$ to the intrinsic system defined by conditions
\begin{equation}\label{a8a}
t_{21}=0\;, \quad\quad t_{22}=t_{22}^{\displaystyle\ast}\;,
\end{equation}
and then diagonalizing the mean field $F$ in the intrinsic
frame. In cases when the $\hat S^+$ pair has the seniority form,
i.e., contains pairs of time-reversed states with equal
amplitudes as is in the single-$j$ shell (\ref{a4a}), we may
further simplify expression (\ref{a8}) by using the fact that
$F$ is a time-even matrix, i.e.,
\begin{equation}\label{a8b}
   C_{mm'}^+=\frac{2 t_0(-1)^{j+m'}}{\sqrt{2j+1}}
                \bigl( e^{2F} \bigr)_{m,-m'}\;.
\end{equation}

\section{Restoration of broken symmetries}
\label{sec6}

It is well known that the GCM for the gauge angle and the Euler
angles used as generator coordinates is equivalent to the exact
particle-number and angular-momentum projections \cite{LB68,Won75,RS80}.
In the present study we use this property to restore those broken
symmetries exactly, and then the GCM equation is solved in the intrinsic
frame of reference.

\subsection{Particle-number projection}
\label{sec2}

A standard way \cite{Bay60,DMP64,Mac70,RS80} to perform the particle-number
projection is to introduce a gauge angle and to integrate over
its $(0,2\pi)$ domain. Using the generating function in the form
(\ref{a6}), the projected state is
\widetext
\begin{equation} \label{p1}
   \mid\!\!Nt_2\rangle\equiv
   \hat P_N |t_0t_2\rangle = {1\over 2\pi} \!\int_0^{2\pi}\!\! d\phi
                      \,\,e^{-iN\phi}
   \exp\biggl(\case{1}{2} \sum_{mm'} C_{mm'}^+ e^{2i\phi}
                    a_m^+ a_{m'}^+ \biggr) |0\rangle \;.
\end{equation}
Such an integration leaves only the components of the wave
function with the particle number $N$. This is based on the
following orthogonality relation valid for any integer $N$ and
$N'$
\begin{equation} \label{p5}
  {1\over 2\pi} \!\int_0^{2\pi}\!\! d\phi\,\,
                  e^{i(N'-N)\phi} = \delta_{NN'}\;.
\end{equation}
When calculating matrix elements of any arbitrary
particle-number-conserving operator $\hat O$ between the Thouless states,
the integration over the gauge angle can be done only for one of
those states
\begin{equation} \label{p10}
    \langle N t_2|\hat O|N t_2'\rangle\equiv
    \langle t_0t_2|\hat O \hat P_N |t_0't_2'\rangle =
          {1\over 2\pi} \!\int_0^{2\pi}\!\! d\phi\,\,
         e^{-iN\phi}  \langle t_0t_2|\hat O |e^{2i\phi}t_0',t_2'\rangle \;,
\end{equation}
which amounts to multiplying the Thouless matrix $C'$
(or, in other words, the coefficient $t_0'$)
of the ket state by the factor $e^{2i\phi}$ and integrating over $\phi$.

When there is an upper limit $N_{\text{max}}$ on the particle
number of the system, as is the case in the single-$j$ shell,
the integral (\ref{p10}) can be discretized as proposed
by Fomenko \cite{Fom70,Sor72}.  Instead of Eq.{\ }(\ref{p5})
one can write
\begin{equation} \label{p15}
    {1\over N_{\text{max}}+1} \sum_{n=0}^{N_{\text{max}}}
               e^{i(N'-N)n\triangle\phi} =
    \left\{
          \begin{array}{cl}
             1     &  \quad\text{for}\quad N'=N \\
             {\displaystyle
             {1\over N_{\text{max}}+1}
             \frac{1-\exp\left[{i(N'-N)(N_{\text{max}}+1)
                                        \triangle\phi}\right]}
                  {1-\exp\left[{i(N'-N)\triangle\phi}\right]}
             }
                   &  \quad\text{for}\quad  N'\ne N
          \end{array}
    \right. \;.
\end{equation}
Taking $\triangle\phi$=$2\pi/(N_{\text{max}}$+1), the numerator of
the expression for $N'$$\ne$$N$ vanishes for all $N$ and $N'$,
whereas the denominator does not vanish for any $N$ and $N'$ not
larger than $N_{\text{max}}$, i.e.,
\begin{equation} \label{p15a}
    {1\over N_{\text{max}}+1} \sum_{n=0}^{N_{\text{max}}}
             e^{i(N'-N)n\triangle\phi} =
             \delta_{NN'} \quad\text{for}\quad N,N'\leq N_{\text{max}}\;.
\end{equation}

For  systems with even particle numbers we may repeat a similar
discretization using only $\case{1}{2}N_{\text{max}}$ points and
$\triangle'\phi$=$\pi/(\case{1}{2}N_{\text{max}}$+1), which gives
\begin{equation} \label{p20}
    {1\over \case{1}{2}N_{\text{max}}+1}
         \sum_{n=0}^{\case{1}{2}N_{\text{max}}}
         e^{i(N'-N)n\triangle'\phi} =
         \delta_{NN'} \quad\text{for even}\quad N,N'\leq N_{\text{max}}\;.
\end{equation}
Furthermore, assuming that $\case{1}{2}N_{\text{max}}$ is also
even, or proceeding as if $N_{\text{max}}$ was larger by 2 when
$\case{1}{2}N_{\text{max}}$ is odd, the left hand side of
Eq.{\ }(\ref{p20}) can be rewritten in a form of the sum over
$n$ from $-\case{1}{4}N_{\text{max}}$ to
$\case{1}{4}N_{\text{max}}$, and finally
\begin{equation} \label{p25}
   {2\over \case{1}{2}N_{\text{max}}+1} \biggl[ {1\over 2} +
      \Re \sum_{n=1}^{\case{1}{4}N_{\text{max}}}
             e^{i(N'-N)n\triangle'\phi}\biggr] =
          \delta_{NN'} \quad\text{for even}\quad N,N'\leq N_{\text{max}}\;.
\end{equation}
Since the unprojected GCM states are time-even, the matrix
elements of time-even operators are real and then the
prescription for the projected matrix elements takes the
following form
\begin{equation} \label{p30}
   \langle N t_2 | \hat O | N t_2' \rangle =
   {2\over \case{1}{2}N_{\text{max}} + 1} \biggl[ {1\over 2}
   \langle t_0 t_2 | \hat O | t_0' t_2' \rangle +
   \Re \sum_{n=1}^{\case{1}{4}N_{\text{max}}} e^{-iNn\triangle'\phi}
  \langle t_0 t_2 | \hat O | e^{2in\triangle'\phi}t_0',t_2' \rangle\biggr]\;,
\end{equation}
where only $\case{1}{4}N_{\text{max}}$ integration points are required.

\narrowtext
The standard formulas resulting from the Wick \cite{Wic50,RS80}
and Thouless \cite{Tho60,RS80} theorems applied to evaluate
$\langle t_0t_2 | \hat O | e^{2in\triangle'\phi}t_0',t_2' \rangle$
matrix elements contain the products of matrices $C^+ C'$,
cf.{\ }Eq.{\ }(\ref{a6}).  Since using the projection method
requires many such matrix multiplications a very efficient
algorithm can be obtained by transforming $C^+ C'$ to the basis
in which it is diagonal (Appendix \ref{AppA}). Then the matrix
multiplications reduce to simple sums and products of the
corresponding eigenvalues.  At the same time the sign problem of
the GCM kernels is avoided by considering only one
half of pairwise degenerate eigenvalues \cite{NW83} .

\subsection{Angular-momentum projection}
\label{sec1c}

The practical techniques of the exact angular momentum projection
were discussed many times in the literature \cite{SGF84,BH84}.
We review here the formalism in the form suitable for the description
of the quadrupole degrees of freedom.

Let us denote by $\mid$$N\beta\gamma\rangle$ the
particle-number-projected states (\ref{p1}) in the intrinsic
frame defined by Eq.{\ }(\ref{a8a}), where $\beta$ and
$\gamma$ are the axial coordinates on the
$t_{20}$--$t_{22}$ plane
\begin{equation}\label{g0}
t_{20}=\beta\cos\gamma\;,\quad\quad \sqrt{2}t_{22}=\beta\sin\gamma\;.
\end{equation}
According to standard prescriptions \cite{RS80,RW54,PY57,LB68,Mac70},
the angular-momentum-projected GCM state $|nNIM\rangle$ has the
form
\begin{equation}\label{g1a}
   \mid\!nNIM\rangle = \sum_{K\ge 0}
                        \!\int\!\! d\beta d\gamma \;
              \frac{g_{nIK}(\beta,\gamma)}{1+\delta_{K0}}
              \mid\!\!NIMK\beta\gamma\rangle \;,
\end{equation}
where the angular-momentum-projected intrinsic state
$|NIMK\beta\gamma\rangle$ is given by
\widetext
\begin{equation} \label{g1}
   |NIMK\beta\gamma\rangle = {2I+1\over 8\pi^2}
              \!\int\!\! d\Omega\,
              \biggl[ D_{MK}^{I^{\displaystyle\ast}}(\Omega)
           + (-1)^I D_{M,-K}^{I^{\displaystyle\ast}}(\Omega)\biggr]
              \hat R(\Omega)|N\beta\gamma\rangle \;.
\end{equation}
The rotation operator $\hat R(\Omega)$ depends on the three
Euler angles, $\Omega$$\equiv$$\phi,\theta,\psi$.  Due to the $D_2$
symmetry of the GCM states (\ref{a3}) the sum in
Eq.{\ }(\ref{g1a})  is restricted to non-negative even values
of $K$.  Using states (\ref{g1a}) one obtains the GCM
(Hill-Wheeler) equation in the form \cite{HW53,OY66,Cor71}
\begin{equation} \label{g5}
  \sum_{K'\ge0} \!\int\!\! d\beta' d\gamma'\,
  \biggl[{\cal H}_{KK'}^I(\beta,\gamma;\beta',\gamma')-E^I_n
        {\cal N}_{KK'}^I(\beta,\gamma;\beta',\gamma')\biggr]
        g_{nIK}(\beta,\gamma) = 0 \;,
\end{equation}
where
${\cal H}$ and ${\cal N}$ are the GCM kernel matrices of the
hamiltonian and of the norm, respectively, defined as
\begin{eqnarray} \label{g10}
  {\cal O}_{KK'}^I (\beta,\gamma;\beta',\gamma')
     &=& {2I+1\over 8\pi^2} \triangle_{KK'}
         \!\int\!\! d\Omega \;
         \langle N\beta\gamma|\hat O \hat R(\Omega)|N\beta'\gamma'\rangle
                                                            \\ \nonumber
     &\times& \biggl[D_{KK'}^{I^{\displaystyle\ast}}(\Omega)
                   + D_{-K-K'}^{I^{\displaystyle\ast}}(\Omega)
            + (-1)^I D_{K-K'}^{I^{\displaystyle\ast}}(\Omega)
            + (-1)^I D_{-KK'}^{I^{\displaystyle\ast}}(\Omega)\biggr] \;,
\end{eqnarray}
and $\triangle_{KK'}$ stands for
$1/((1$+$\delta_{K0})(1$+$\delta_{K'0}))$.  Since evaluation of
kernels in Eq.{\ }(\ref{g10}) requires three dimensional
integration, this is the most time consuming part of the
numerical calculation.

Because all indices $K$ and $K'$ are even, we may use the
explicit form \cite{Var85} of the Wigner matrices,
$D_{KK'}^I(\Omega)$=$e^{iK\phi}d_{KK'}^I(\theta)e^{iK'\psi}$ to
rewrite (\ref{g10}) in the following way
\begin{eqnarray} \label{g15}
  {\cal O}_{KK'}^I (\beta,\gamma;\beta',\gamma')
   &=& 2{2I+1\over 8\pi^2} \triangle_{KK'}
     \!\int_0^{2\pi}\!\! d\phi \!\int_0^\pi\!\! d\theta \sin\theta
     \!\int_{0}^{2\pi}\!\! d\psi \;
         \langle N\beta\gamma|\hat O \hat R(\phi,\theta,\psi)|
                 N\beta'\gamma'\rangle
                                                              \nonumber \\
   &\times&  \biggl[ d_{KK'}^I(\theta) \cos(K\phi+K'\psi)
            + (-1)^I d_{K-K'}^I(\theta) \cos(K\phi-K'\psi) \biggr] \;.
\end{eqnarray}\narrowtext\noindent%
By the same token the integration intervals over $\phi$ and
$\psi$ can be reduced from $(0,2\pi)$ to $(0,\pi)$.  Moreover,
the Wigner functions $d_{KK'}^I(\theta)$ depend then on
$x$=$\cos\theta$ only, and the integral $\displaystyle\int_0^\pi\!\!
d\theta \sin\theta$ can be rewritten as
$\displaystyle\int_{-1}^1\!\!dx$.  Furthermore, symmetry properties
\cite{Var85} of $d_{KK'}^I(\theta)$ allow also to reduce
integration over $x$ to the interval (0,1).  Finally, the
integration domain in (\ref{g15}) is eight times smaller and the
factor 2 in front is replaced by 16.

Numerical integration over $\phi$ and $\psi$ Euler angles is
done by means of the Tchebyshev method \cite{Abr65}, i.e., values
of the integrand in equally spaced points are summed up with
equal weights.  This method is exact when number of points is
taken to be $I_{\text{max}}$+1, where $I_{\text{max}}$ is the
maximum spin of the good-angular-momentum components in the
intrinsic state $|N\beta\gamma\rangle$.  Integration over $x$ is
done using the Gauss-Legendre method \cite{Abr65} applied to the
(0,1) interval.  Again, exact result is obtained by setting the
number of points to $I_{\text{max}}$+1.

\section{GCM calculations in the intrinsic frame}
\label{sec3}

The exact particle-number and angular-momentum projections allow
us to calculate hamiltonian and norm kernels, ${\cal
H}_{KK'}^I(\beta,\gamma;\beta',\gamma')$ and ${\cal
N}_{KK'}^I(\beta,\gamma;\beta',\gamma')$, for arbitrary pairs of
intrinsic deformations $(\beta,\gamma)$ and $(\beta',\gamma')$, which
define the GCM equation (\ref{g5}). By a discretization of the
$\beta$ and $\gamma$ variables this integral eigen-equation is
transformed into a matrix eigen-equation, for which the kernels have
to be calculated between all pairs of selected points
($\beta_n,\gamma_n$), $n$=1,$\dots$,$N_d$.

In principle, the invariant measure
$\beta^4|\sin3\gamma|d\beta{}d\gamma$ should be used in the GCM
equation (\ref{g5}) for quadrupole motion.  However, since we
discretize this integral equation, the integration measure
is immaterial. Moreover, we may then also include points for
axial ($\gamma$=0) and spherical ($\beta$=0) shapes, which in
the discretized GCM equation represent a certain volume of the
phase space around points for which the invariant measure
vanishes.

For a given $j$, the main factor determining the total computing
time is the number $N_d$ of the mesh points ($\beta_n,\gamma_n$)
on the $\beta$--$\gamma$ plane, because the number of projected
kernels to be calculated is equal  $\case{1}{2} N_d(N_d+1)$.  Each of
these kernels requires calculating $(I_{\text{max}}$+$1)^3$
overlaps for different Euler angles, Eq.{\ }(\ref{g15}).  In
Fig.{\ }\ref{f1} we present the CPU times required to
calculate one hundred of such overlaps as a function of the
single-particle angular momentum $j$ of the shell.
Full circles denote results obtained on a 50MHz PC-486 scalar computer.
This computing time grows roughly as a third-order polynomial in $j$
(solid line),
which is a much slower
increase than what would be required for an exact solution.
The times obtained on a vector supercomputer CRAY Y/MP EL98 are marked
with full squares.
They were multiplied by a factor of 20 before plotting them
together with the PC results.
The increase of time with $j$ is here much slower, which is due to
a better performance of the vector processor obtained for
larger matrices. In fact, this time can be approximated by
a polynomial of a second degree in $j$ (solid line).

In the present study we have performed the GCM calculations for
$j$=31/2 and for $j$=15/2. For the latter value, the exact
results can easily be obtained and were used to test the GCM method.

As discussed in Sec.{\ }\ref{sec1b}, the average value of
angular momentum in the intrinsic states $|N\beta\gamma\rangle$
increases with $\beta$. Similarly, for large values of $I$ the
collective weight functions $g_{nIK}(\beta,\gamma)$ have large
components at large $\beta$. Choosing the mesh of points
($\beta_n,\gamma_n$) with $\beta_n$ smaller than a given value
$\beta_{\text{max}}$ is equivalent to restricting the
variational space to the values of spins smaller than a certain
maximum value $I_{\text{max}}$.  In calculations we used
$\beta_{\text{max}}$=6 which allows to properly describe spins
up to $I_{\text{max}}$=18.

Two meshes of points ($\beta_n,\gamma_n$) were used. The first
one was composed of 10 points uniformly distributed in the
sector
$\beta$$\leq$$\beta_{\text{max}}$, 0$\leq$$\gamma$$\leq$$60^\circ$
(Fig.{\ }\ref{f2}, full circles) in such a way that overlaps
of the neighboring points do not differ too much one from
another.  The average overlap turns out to be $0.73$ for $N$=10
particles in the $j$=31/2 shell and $0.80$ for $N$=8 and
$j$=15/2.  To test the influence of the number of points on the
results, the second mesh comprising $19$ points was constructed
by adding one point in the middle of each triangle formed by
three neighboring points of the 10-point lattice.

In Fig.{\ }\ref{f2} the PES in the
intrinsic frame of reference is presented for the pure
quadrupole force.  The upper part shows the results for $N$=8
and $j=15/2$, in which case the PES has a deformed
$\gamma$-unstable valley. Two absolute minima are at
$\beta$$\sim$3.5 for $\gamma$=$0^\circ$ and $\gamma$=$60^\circ$
with a low saddle in between.  The lower part of
Fig.{\ }\ref{f2} shows the PES for $N$=10 and $j$=31/2.  In
this case there appears a strongly pronounced oblate valley that
extends towards large values of $\beta$.

The quadrupole moment of the intrinsic state can be calculated
as an expectation value of the quadrupole operator in that
state,
\begin{equation}\label{i2}
    Q_{\mu}(\beta,\gamma)
        =\langle N\beta\gamma|\hat Q_{\mu}|N\beta\gamma\rangle \;.
\end{equation}
The intrinsic frames of reference for the deformation tensor
$t_2$ and for the quadrupole tensor $Q$ coincide, i.e., the only
non-zero components of $Q$ are
\begin{equation}\label{i3}
   Q_0=Q\cos\delta \quad\mbox{and}\quad \sqrt{2}Q_2=Q\sin\delta \;,
\end{equation}
which defines the radial coordinates in the $Q$--$\delta$ plane,
in analogy to Eq.{\ }(\ref{g0}).  When the mesh points of
Fig.{\ }\ref{f2} are plotted in the $Q$--$\delta$ variables
they appear to be distributed in a more uniform way than in the
$\beta$--$\gamma$ variables.  On the other hand, the topology of
the PES is the same in both representations, with one important
difference, namely, due to a finite dimension of the
single-$j$-shell space, the quadrupole moment $Q$ has its upper
limit $Q_{\text{max}}$.  In Table~\ref{t3} the values of
$Q_{\text{max}}$ are shown for several particle numbers in the
$j$=15/2 and $j$=31/2 shells.

For $j=$31/2 the intrinsic states $|N\beta\gamma\rangle$ with
the $\beta$ coordinates beyond $\beta$$\approx$10 all have quadrupole
moments close to the maximum value $Q_{\text{max}}$.  This
means that there is no reason to extend the generating
coordinates outside the $\beta_{\text{max}}$=10 circle, and
in fact the value of $\beta_{\text{max}}$=6 used in the
calculations is sufficient.

\subsection{GCM solutions for $j$=15/2 and $N$=8}
\label{sec4b}

In this section we present the GCM calculations for
$j$=15/2 and $N$=8, where the exact solutions are available
and can be used to test the properties of the GCM approach.
The Hill-Wheeler equation (\ref{g5}) is solved by using
the standard method of calculating the square root of the overlap matrix
${\cal N}^I_{KK'}$ described in Ref.{\ }\cite{RS80}.
In what follows we compare the results obtained for 10
and 19 mesh points of intrinsic deformations, as defined
in Sec.{\ }\ref{sec3}.

As is well known, the GCM overlap matrix has the spectrum with the
zero value as an accumulating point,
cf.{\ }the discussion in Ref.{\ }\cite{BDF90}. In principle
this matrix is strictly positive definite but in
practical calculations the negative eigenvalues may appear at the
level of the numerical precision. This may be related to a finite
precision of the diagonalization method, and especially to the numerical
noise which may appear in calculating the elements of the overlap matrix.
In Figs.{\ }\ref{f15}(a) and \ref{f15}(b)
we present the absolute values of the overlap
matrix for $I$=0 and $I$=6, respectively, obtained with 19 mesh points.
For $I$=0 the exact spectrum contains 7 states,
see Fig.{\ }\ref{f14} or Table~\ref{t2}, and the overlap
matrix has the same number of large eigenvalues separated
by 10 orders of magnitude from the remaining 12 eigenvalues
which represent the numerical noise. This illustrates a rare
situation when the GCM generating functions exhaust the
corresponding Hilbert space completely.

A more typical spectrum of the overlap matrix appears for $I$=6,
where the total number of eigenvalues equals to 76
(its dimension equals the product of the number of points (19)
and of the number of $K$ components (4), $K$=0,2,4, and 6).
Even if this dimension is larger then the number of the exact
$I$=6 states (31), the GCM generating functions do not exhaust the
Hilbert space and no gap in the overlap spectrum occurs.
In calculating the square root of the overlap one has to
decide how many norm eigenvalues should be retained.
Since the negative eigenvalues reflect the appearance of the numerical
noise, their magnitude can be used as an indication of how much
the results are perturbed by numerical uncertainties.
Therefore, in the following we shall use and discuss a useful
parameter denoted by $k_{\mbox{\scriptsize crit}}$,
which is equal to the number
of norm eigenvalues larger than the absolute value of the smallest
eigenvalue. In Fig.{\ }\ref{f15} we present a graphic
illustration of how this parameter is defined.

Let us denote by $k$ the number of norm
eigenvalues retained in solving the GCM equation.
A choice of a proper value of $k$ should
be based on studying the stability
of solutions when this parameter increases.
In Fig.{\ }\ref{f3} we show the GCM energies for $I$=0 (full circles)
obtained by using 10 (a) and 19 (b) mesh points.
The exact energies are also shown (open squares) and placed
between the results obtained with $k$=$k_{\mbox{\scriptsize crit}}$
and $k$=$k_{\mbox{\scriptsize crit}}$+1.
It can be seen that the GCM calculation
with $k$=7 give the exact spectrum with a very good
accuracy. This is due to the fact that that the complete
Hilbert space is obtained in this case. For values of $k$=1,2, or 3
one obtains a fair description of 1,2, or 3 lowest states. On the
other hand, the fourth state which first appears for $k$=4
is correctly split into two exact solutions only after all the 7 basis states
are included. We may see this kind of effect in many other
examples and one may interpret it as a manifestation
of the coupling of collective and non-collective degrees of freedom.
Within this interpretation, for $k$=4 the GCM correctly singles out the
fourth collective state, which in reality is fragmented among
the forth and the fifth exact states.

As soon as we include in the basis more states than the critical
value $k_{\mbox{\scriptsize crit}}$,
there appear in the spectrum spurious states
which have nothing in common with the exact solutions.
Such states appear in an unpredictable way at various energies
and in realistic calculations cannot be distinguished from
the real ones. In fact, they also pollute the wave functions of
states which have energies seemingly stable when $k$
increases beyond the value of $k_{\mbox{\scriptsize crit}}$.
Therefore, one should
always try to avoid using too large values of $k$ and
an identification of a highest acceptable value is essential
\cite{spur}.

In Fig.{\ }\ref{f4} we present similar results for the $I$=2 states.
For both mesh sizes we
obtain a good description of the four lowest states by using
$k$=4. At higher energy one may identify two other structures
which appear at $k$=5 and $k$=9, and
for increasing $k$ become fragmented.
For the 10-point calculation,
except of the four lowest states,
none of the higher excited states is reproduced with
$k$$\leq$$k_{\mbox{\scriptsize crit}}$=12.
On the other hand, for the 19-point mesh
the exact spectrum is obtained for
$k$$\leq$$k_{\mbox{\scriptsize crit}}$=16, because the full Hilbert space
is then again exhausted by the GCM generating functions.
It is interesting to note that for  10  mesh  points  the  fifth  exact  state
appears
only beyond the value of $k$=$k_{\mbox{\scriptsize crit}}$.
This may illustrate
an approximate character of our prescription to define the
critical value $k_{\mbox{\scriptsize crit}}$. For 19 points the fifth state
is stable already at $k$=$k_{\mbox{\scriptsize crit}}$$-$2.

Based on this results we will suppose in what follows
that using a relatively small number of mesh points of the
intrinsic deformations
(as compared to the number of exact states),
one obtains a fair description of the collective subspace,
and that only states
with strong collective components are then present in the GCM spectra.
By increasing the
number of mesh points one includes noncollective configurations
and the GCM
yields collective states mixed with the noncollective ones.
A separation of states into the collective and noncollective
subspace is of course not a well defined procedure (both in the
experiment and in the exactly solvable model discussed here).
Which states are called collective is therefore a subject of a
model interpretation.
With increasing excitation energy the collectivity is gradually lost
and the mixing with noncollective states increases. We may see, however,
that the GCM can be used
as a practical tool to select and decouple
the collective space.

In  Fig.{\ }\ref{f7} we compare the GCM spectra for spins up to $I$=6
with the exact results. Again, both the 10-point
(a) and 19-point (b) results are shown.
In the GCM calculations the cut-off parameters $k$ were
chosen to be equal to the critical values $k_{\mbox{\scriptsize crit}}$,
which are listed in Table~\ref{t2} together
with the corresponding numbers of exact states for a given spin.
In all cases we have checked that the
GCM states have correct average values of spin
$\langle\hat{I}^2\rangle$ and particle numbers,
$\langle\hat{N}\rangle$ and $\langle\hat{N}^2\rangle$,
and that the average values of the hamiltonian
$\langle\hat{H}\rangle$ are equal to the
GCM eigenenergies.
These turn out to be useful criteria of identifying spurious states,
because as soon as the spurious states start mixing with the physical ones
the above average values become perturbed, even if
the symmetries have been exactly restored when calculating the GCM kernels.

In the GCM spectra there is no
$1^+$ states because they belong to
representations of the point group $D_{2}$ which are not
included in the present calculations.
Our choice of the generating functions (\ref{a3}) restricts them
to the symmetric representation (+++) \cite{BM75},
and therefore such is the symmetry
of all GCM eigenstates, whereas the $1^+$ states belong to mixed-symmetry
representations.
In principle, some other high-spin many-particle states in the single-$j$
shell
may also belong to the mixed-symmetry representations and will not
be accessible in our implementation of the GCM. This is not a case
for the $N$=8 and $I$=2 states in the $j$=15/2 shell discussed here
because the GCM reproduces these states exactly. On the other
hand, the GCM method gives much less odd-spin states then even-spin
states as compared to the exact solutions, Fig.{\ }\ref{f7}.
This can be understood by recalling that, for example,
in the asymmetric rotor model
\cite{BM75} only ($I$$-$1)/2 odd-spin states
belong to the symmetric representation as compared to the
$I$/2$+$1 even-spin
states. One should also stress that the standard Bohr collective quadrupole
model \cite{RS80} describes the symmetric states only.

Comparing the results for 10 and 19 points, Fig.{\ }\ref{f7},
we may see that only three collective $I$=3 states are found
in our GCM space in the former case. The lowest one clearly reproduces
the lowest exact state while the first-excited state is probably
fragmented among three excited exact solutions.
For 19 points 7 states are found of which the lowest 4 correspond
to the exact ones.
For $I$=5 only two collective states are found; the third excited
state appears for 19 points at the correct energy, but it is absent
in the 10 point results. Similar analysis shows that the four
lowest $I$=4 and $I$=6 states can be considered as collective ones.

The lowest GCM states with spins up to $I$=18 are shown
in  Fig.{\ }\ref{f8}.
Full circles denote states with positive signature $(-1)^I$
and open circles with the negative signature.
The states are connected by lines according to their
stretched E2 matrix elements, see Sec.{\ }\ref{sec4a}.
In order to illustrate the convergence properties of the
GCM we draw such lines only between states which have the GCM
absolute energies precise to better than 3.5\% as compared with
the exact values. Again we see that the 10-point GCM calculation
selects two lowest collective bands
while in the 19-point results
many higher excited, and not necessarily collective states are
found.

\subsection{GCM solutions for $j$=31/2}
\label{sec4c}

In Fig.{\ }\ref{f3bis} we
show the GCM energies for $I$=0 states of $N$=8 particles
obtained for different
values of the cut-off parameter $k$.
The vertical dashed lines are placed
between the results obtained with $k$=$k_{\mbox{\scriptsize crit}}$
and $k$=$k_{\mbox{\scriptsize crit}}$+1
to show the values of the critical cut-off parameters
$k_{\mbox{\scriptsize crit}}$
determined as described in Sec.{\ }\ref{sec4b}.
For the 10-point mesh in the intrinsic frame of reference,
Fig.{\ }\ref{f3bis}(a), we obtain the two lowest $I$=0 states
for $k$=1 and 2, respectively. Only the first of them has
a fairly correct energy already for $k$=1; the energy of the
second one decreases rapidly and stabilizes around $k$=6.
According to our method of identifying collective states,
the third and probably the fourth collective $I$=0 states appear at
$k$=3 and 5, respectively, and then become strongly fragmented.
This picture is confirmed by the 19-point results presented in
Fig.{\ }\ref{f3bis}(b).

A similar analysis of the $I$=2 states,
Fig.{\ }\ref{f4bis}, shows that the lowest
three of them have a collective nature while other
three collective structures
(appearing at $k$=4, 8, and 11 in
Fig.{\ }\ref{f4bis}(b)) become completely mixed and disappear in
the dense non-collective spectrum.

Figs.{\ }\ref{f9}(a--f) show the GCM spectra
for particle numbers $N$=4 to $N$=14, respectively,
obtained by using the 19-point mesh of intrinsic deformations.
All presented GCM eigenenergies correspond to states
fulfilling the criterion of stability with respect
to the cut-off parameters $k$$\leq$$k_{\mbox{\scriptsize crit}}$ as well
as have correct average values of
$\langle\hat{I}^2\rangle$,
$\langle\hat{N}\rangle$, $\langle\hat{N}^2\rangle$, and
$\langle\hat{H}\rangle$, as discussed in Sec.{\ }\ref{sec4b}.

The spectrum for $N$=4 can be compared with the exact results
shown in Fig.{\ }\ref{f10}. It is seen that the lowest GCM collective
bands agree very well with the exact calculations.
This is due to the fact that the dimension of the Hilbert space is here
not too large and the GCM generating functions seem to exhaust it,
at least for lower spins.
The states for spins higher than $I$=12 are
reproduced less accurately and the bands gradually disappear,
even if in the exact results they can be followed up to
still higher spins.
The bending up at higher spins of the $n_{\gamma}$=2 GCM bands
illustrates the effect of decreasing possibility of the GCM states
to exhaust the exact Hilbert space.

For all particle numbers one may clearly distinguish
at least five lowest collective bands. All of them are
similar to those for $N$=4 and may accordingly be interpreted as
the $\gamma$ vibrations coupled to an axial rotor,
see Sec.{\ }\ref{sec4a}.
In all these cases one cannot see any candidate for
low-energy $\beta$ vibrations \cite{CB94}.

With an increasing particle number the GCM
collective quadrupole bands loose their regularity. This effect
can probably be attributed
to a decreasing ability of the generating functions used in this study
to exhaust the space of quadrupole collective excitations.
But even for $N$=14 the low-energy collective bands are visible.
On the other hand, for the $\gamma$-unstable case corresponding to
the half-filled shell
($N$=16), we could not obtain any stable GCM solutions
even by using the 19-point mesh of generating coordinates.

\section{Summary and conclusions}
\label{sec5}

In the present paper  we have discussed  properties of
states of particles moving in a single-$j$ shell and interacting
with the quadrupole-quadrupole force. This system exhibits
collective as well as non-collective features and we
have analyzed a possibility to  select the collective
subspace by using the generator coordinate method (GCM).

The GCM generating functions have been constructed
as coherent excitations built from the single-particle
quadrupole operator. This type of  building blocks has
been proposed in the single-particle coherent excitation model
(SCEM) which constitutes an interesting
alternative to other models based on a building-block concept,
e.g., the bosonic IBM model \cite{AI75}
and the algebraic Ginocchio models \cite{Gin78}.  A collective model
constructed in this way is based on purely fermionic objects
and requires no closed algebraic properties.

Depending on the number of different intrinsic-state configurations
used in the GCM we obtain different precision of results as
compared with exact solutions. When this number is
relatively small, the GCM is shown to correctly select the
quadrupole collective structures, while when this number is large
all exact states are obtained with high precision.

We have shown that in the single-$j$ shell
the low-lying quadrupole collective states can be interpreted in
the frame of a standard geometrical collective model,
either in its limit of an axial rotor coupled to $\gamma$-vibrations
or in the limit of a $\gamma$-unstable motion.
In the single-$j$ shell we did not find collective bands of the
beta-vibrational character, even for $j$ as large as 31/2.

This research was supported in part by the
Polish State Committee for Scientific Research under Contract
No. 20450~91~01. Numerical calculations were performed
at The Pittsburgh Super Computer Center
under grant No. PHY900027P
and at
The Interdisciplinary Centre for Mathematical and Computational Modeling
(ICM) of Warsaw University.

\appendix
\section{Canonical basis for transition matrix elements}
\label{AppA}

Canonical basis for antisymmetric matrices has been introduced
in Refs.{\ }\cite{Zum62,BM62} and, together with the Wick
\cite{Wic50,RS80} and
Onishi \cite{Low55,OY66,RS80} theorems, can be used to facilitate
calculation of average values of fermion operators in
quasiparticle states.  Here we extend the concept of the
canonical basis to non-diagonal (transition) matrix elements.

We begin by considering the eigen-equation for the product of
two complex antisymmetric matrices  $C^+$ and $C'$
   \begin{equation}\label{Ap1}
   \left(C^+C'\right)W=WD\;,
   \end{equation}
where the columns of the matrix $W$ are linearly independent
vectors of the Jordan basis, and $D$ is the Jordan block matrix
with the eigenvalues $D_i$ at the main diagonal,
the zeros or ones just above the main diagonal, and zeros elsewhere.
In general, the eigenvalues $D_i$ are complex.

Multiplying from the left-hand and right-hand sides the
eigen-equation (\ref{Ap1}) by $W^{-1}$ and then transposing, we
obtain that
   \begin{equation}\label{Ap2}
   C'\left(C^+{W^{-1}}^T\right)={W^{-1}}^T D^T\;,
   \end{equation}
which multiplied by $C^+$ from the left-hand side gives
   \begin{equation}\label{Ap2a}
   \left(C^+C'\right)\left(C^+{W^{-1}}^T\right)
     =\left(C^+{W^{-1}}^T\right)D^T\;.
   \end{equation}
We may now discuss several specific cases depending on the
degeneracy of eigenvalues.

\subsection{Non-degenerate eigenvalues}
\label{AppAa}
Suppose that the $i$-th eigenvalue is non-degenerate. Then the
$i$-th column of the matrix $C^+{W^{-1}}^T$ is proportional to the $i$-th
column of $W$,
   \begin{equation}\label{Ap3}
   \left(C^+{W^{-1}}^T\right)_{mi} = \alpha W_{mi}\;,
   \end{equation}
where $\alpha$ is a proportionality constant.
We keep the notation of indices
$m$ and $m'$ denoting the single-particle states, although the
results derived here do not depend on the assumption
that these states belong to the single-$j$ shell.
Multiplying both
sides of Eq.{\ }(\ref{Ap3}) by $W^{-1}_{im}$ and summing over
$m$ we obtain
   \begin{equation}\label{Ap4}
   \alpha = \left(W^{-1}C^+{W^{-1}}^T\right)_{ii}\;,
   \end{equation}
which gives $\alpha$=0, because the right-hand-side is a
diagonal matrix element of an antisymmetric matrix.
{}From Eq.{\ }(\ref{Ap3}) we now see that the
$i$-th column of $C^+{W^{-1}}^T$ is equal to zero and hence from
Eq.{\ }(\ref{Ap2}) the corresponding eigenvalue $D_i$ is equal
to zero. Therefore, there can be at most one
non-degenerate eigenvalue and it must be equal to zero.
All other eigenvalues must be at least pairwise degenerate.

\subsection{Doubly degenerate eigenvalues with one eigenvector}
\label{AppAb}
Suppose now that the matrix $C^+C'$ has a doubly degenerate
eigenvalue $D_i$ for which only one eigenvector exists, and let
this eigenvector be equal to the $i$-th column of $W$. Let the
$\tilde\imath$-th column of $W$ contains its partner in the
2$\times$2 Jordan block corresponding to this doubly degenerate
eigenvalue. From (\ref{Ap1}) we then have
   \begin{mathletters}\begin{eqnarray}
   \left(C^+C'W\right)_{mi} &=& D_iW_{mi}\;, \label{Ap5a} \\
   \left(C^+C'W\right)_{m\tilde\imath} &=& D_iW_{m\tilde\imath}
                                     +     W_{mi} \label{Ap5b}\;,
   \end{eqnarray}\end{mathletters}\noindent%
and from (\ref{Ap3}),
   \begin{mathletters}\begin{eqnarray}
   \left[\left(C^+C'\right)
         \left(C^+{W^{-1}}^T\right)\right]_{m\tilde\imath} &=&
         D_i\left(C^+{W^{-1}}^T\right)_{m\tilde\imath}\;, \label{Ap6a} \\
            \left[\left(C^+C'\right)\left(C^+{W^{-1}}^T\right)\right]_{mi}
     &=& D_i\left(C^+{W^{-1}}^T\right)_{mi} \nonumber \\
     &+&    \left(C^+{W^{-1}}^T\right)_{m\tilde\imath}\;. \label{Ap6b}
   \end{eqnarray}\end{mathletters}\noindent%
Comparing Eqs.{\ }(\ref{Ap5a}) and (\ref{Ap6a}) we see that
the $\tilde\imath$-th column of $C^+{W^{-1}}^T$ must be
proportional to the $i$-th column of $W$,
   \begin{mathletters}\begin{equation}\label{Ap7}
   \left(C^+{W^{-1}}^T\right)_{m\tilde\imath} = \alpha W_{mi}\;,
   \end{equation}
while multiplying Eqs.{\ }(\ref{Ap5b}) by $\alpha$ and
subtracting from (\ref{Ap6b}) we obtain that
   \begin{equation}\label{Ap8}
   \left(C^+{W^{-1}}^T\right)_{mi}
      = \alpha W_{m\tilde\imath} + \beta W_{mi}\;,
   \end{equation}\end{mathletters}\noindent%
where $\alpha$ and $\beta$ are constants. However, summing
Eq.{\ }(\ref{Ap8}) with $W^{-1}_{mi}$ we obtain $\beta$=0.
Similarly summing Eqs.{\ }(\ref{Ap7}) and (\ref{Ap8}) with
$W^{-1}_{mi}$ and $W^{-1}_{m\tilde\imath}$, respectively, we
obtain that $\alpha$ is simultaneously equal to both transposed
matrix elements of an antisymmetric matrix , and hence
$\alpha$=0 too.  Finally, both vectors at the left-hand sides of
Eqs.{\ }(\ref{Ap7}) and (\ref{Ap8}) vanish, which is in
contradiction with Eq.{\ }(\ref{Ap2}), where the $D$ matrix has
an non-diagonal matrix element equal to one.  As a conclusion we
rule out the possibility of one eigenvector existing for a
doubly degenerate eigenvalue. Similar derivation can be
repeated for higher degeneracies and we conclude that the matrix
$D$ is always diagonal, i.e., the number of eigenvectors of the
non-hermitian matrix $C^+C'$ equals to the number of dimensions.

\subsection{Doubly degenerate eigenvalue with two eigenvectors}
\label{AppAc}
Now we have to again consider the case of the double degeneracy
and analyze the structure of the two corresponding eigenvectors.
Let these eigenvectors be equal to the $i$-th and $\tilde\imath$
columns of the matrix $W$. Similarly as in Sec.{\ }\ref{AppAa}
we show that the $i$-th and $\tilde\imath$-th columns of the
matrix $C^+{W^{-1}}^T$ are also eigenvectors with the same
eigenvalue and therefore must be linear combinations of the
type:
   \begin{mathletters}\begin{eqnarray}
   \left(C^+{W^{-1}}^T\right)_{mi}
      &=& \alpha W_{mi} + \beta W_{m\tilde\imath}\;, \label{Ap9a} \\
   \left(C^+{W^{-1}}^T\right)_{m\tilde\imath}
      &=& \gamma W_{mi} + \delta W_{m\tilde\imath}\;, \label{Ap9b}
   \end{eqnarray}\end{mathletters}\noindent%
where $\alpha$, $\beta$, $\gamma$, and $\delta$ are constants.
Summing Eqs.{\ }(\ref{Ap9a}) and (\ref{Ap9b}) with
${W^{-1}}^T_{mi}$ and ${W^{-1}}^T_{m\tilde\imath}$ we obtain
that $\alpha$=0 and $\delta$=0, respectively, while performing
the summation with these factors reversed we obtain that
$\beta$=$-$$\gamma$.

Similar procedure can be repeated for higher degeneracies with
the same result that the eigenvectors of the matrix $C^+C'$ are
grouped in degenerate pairs, and that these pairs transform the
$C^+$ matrix to its canonical form where it has the form of
2$\times$2 antisymmetric blocks with zeros elsewhere.  Therefore, this
canonical form can be written as
   \begin{equation}\label{Ap10}
   \left({W^{-1}} C^+ {W^{-1}}^T\right)_{ji}
     =s_j c_{j}^{\displaystyle\ast}\delta_{j\tilde\imath}\;,
   \end{equation}
where we use the convention of the state $\tilde\imath$ being a
partner of the state $i$, while
$\tilde{\tilde\imath}$$\equiv$$i$.  The numbers
$c_i$=$c_{\tilde\imath}$ are canonical matrix
elements of the matrix $C^+$ while $s_{i}$=$-$$s_{\tilde\imath}$
are the phase factors, $|s_i|^2$=1, depending on the phase
convention chosen for the canonical matrix elements $c_i$ and on
the phases of columns of the matrix $W$.

Comparing Eqs.{\ }(\ref{Ap1}) and (\ref{Ap10}) we see that the
same matrix $W$ simultaneously transforms the second
antisymmetric matrix $C'$ to its canonical form
   \begin{equation}\label{Ap11}
   \left(W^T C' W\right)_{ji}
     =s_i^{\displaystyle\ast} c'_{i}\delta_{j\tilde\imath}\;,
   \end{equation}
where the canonical matrix elements $c'_i$ of $C'$ are related
to the canonical matrix elements $c_i^{\displaystyle\ast}$ of $C^+$ by
   \begin{equation}\label{Ap12}
     c_i^{\displaystyle\ast} c'_i = D_i\;.
   \end{equation}
Since the numbers $c_i$ depend on the normalization of columns
of $W$, we may chose to work with the canonical basis for which
$c_i$=1 and $c'_i$=$D_i$.  However, a better choice is to use
the normalization in which $c_i^{\displaystyle\ast}$ and $c'_i$ are equal:
   \begin{equation}\label{Ap13}
     c_i^{\displaystyle\ast} = c'_i = \sqrt{D_i}\;,
   \end{equation}
with an arbitrary branch of the square root calculated for
complex numbers $D_i$.

\section{Transition matrix elements in the canonical basis}
\label{AppB}
The Wick theorem allows to express non-diagonal
matrix elements of fermion operators between the Thouless
states (\ref{a6}) through three transition density matrices \cite{RS80},
\widetext
   \begin{mathletters}\begin{eqnarray}
   \rho_{mm'}        &\equiv& \frac{\langle C'|a^+_{m'}a_m  |C\rangle}
                                   {\langle C'|              C\rangle}
                            = \left[(1+C^+C')^{-1}C^+C'\right]_{mm'}\;,
                                                \label{Ap14a} \\
   \kappa_{mm'}      &\equiv& \frac{\langle C'|a_{m'}  a_m  |C\rangle}
                                   {\langle C'|              C\rangle}
                            = \left[(1+C^+C')^{-1}C^+  \right]_{mm'}\;,
                                                \label{Ap14b} \\
   {\kappa'}^+_{mm'} &\equiv& \frac{\langle C'|a^+_{m'}a^+_m|C\rangle}
                                         {\langle C'|        C\rangle}
                            = \left[C'(1+C^+C')^{-1}   \right]_{mm'}\;.
                                                      \label{Ap14c}
   \end{eqnarray}\end{mathletters}\narrowtext\noindent%
Using the canonical forms of the Thouless matrices $C^+$ and $C'$,
Eqs.{\ }(\ref{Ap10}) and (\ref{Ap11}), we obtain the transition
density matrices in the canonical basis:
   \begin{mathletters}\begin{eqnarray}
   \bar\rho_{ji}        &\equiv& \left(W^{-1}\rho W\right)_{ji}
                            = \frac{D_j\delta_{ji}}{1+D_j}\;,
                                                \label{Ap15a} \\
   \bar\kappa_{ji}      &\equiv& \left(W^{-1}\kappa{W^{-1}}^T\right)_{ji}
               = \frac{s_j\sqrt{D_j}\delta_{j\tilde\imath}}{1+D_j}\;,
                                                \label{Ap15b} \\
   {\mbox{${\bar\kappa}'$}}^+_{ji} &\equiv& \left(W^T{\kappa'}^+ W\right)_{ji}
               = \frac{s_i^{\displaystyle\ast}
                 \sqrt{D_i}\delta_{j\tilde\imath}}{1+D_i}\;,
                                                \label{Ap15c}
   \end{eqnarray}\end{mathletters}\noindent%
where the normalization (\ref{Ap13}) was used. We see that in the
canonical basis the pairing-tensor densities are related by
$\bar\kappa$=$-$${\mbox{${\bar\kappa}'$}}^+$, provided phases are
chosen in such a way that the phase factors $s_i$ are real.

We may now express transition matrix elements of fermion operators
through their single-particle matrix elements in the canonical basis.
For the one-body and the pair-transfer operators
   \begin{mathletters}\begin{eqnarray}
     \hat Q   &=& \sum_{mm'} Q_{mm'}  a^+_m a_{m'}
                                                \label{Ap16}\;, \\
     \hat Q^+ &=& \sum_{mm'} Q^+_{mm'}a^+_m a_{m'}
                                                \label{Ap16a}\;, \\
     \hat P^+ &=& \sum_{mm'} P^+_{mm'}a^+_m a^+_{m'}
                                                \label{Ap16b}\;, \\
     \hat P   &=& \sum_{mm'} P_{mm'}  a_m   a_{m'}
                                                \label{Ap16c}\;,
   \end{eqnarray}\end{mathletters}\noindent%
we have
   \begin{mathletters}\begin{eqnarray}
    \frac{\langle C'|\hat Q|C\rangle}
         {\langle C'|       C\rangle}
              &=& \sum_{mm'} Q_{mm'}  \rho_{m'm}
               =  \sum_i \frac{{\bar Q}'_{ii}D_i}{1+D_i}
                                                \label{Ap17}\;, \\
    \frac{\langle C'|\hat Q^+|C\rangle}
         {\langle C'|       C\rangle}
              &=& \sum_{mm'} Q^+_{mm'}  \rho_{m'm}
               =  \sum_i \frac{{\bar Q}^+_{ii}D_i}{1+D_i}
                                                \label{Ap17a}\;, \\
     \frac{\langle C'|\hat P^+|C\rangle}
          {\langle C'|         C\rangle}
              &=& \sum_{mm'} P^+_{mm'} {\kappa'}^+_{m'm}
               =  \sum_i \frac{{\bar P}^+_{i\tilde\imath}
                         s_i^{\displaystyle\ast}\sqrt{D_i}}{1+D_i}
                                                \label{Ap17b}\;, \\
     \frac{\langle C'|\hat P  |C\rangle}
          {\langle C'|         C\rangle}
              &=& \sum_{mm'} P_{mm'}    \kappa_{m'm}
               =  \sum_i \frac{{\bar P}'_{\tilde\imath i}
                         s_i\sqrt{D_i}}{1+D_i}
                                                \label{Ap17c}\;,
   \end{eqnarray}\end{mathletters}\noindent%
where only diagonal or paired matrix elements of matrices
   \begin{mathletters}\begin{eqnarray}
            {\bar Q}'  &=& W^{-1} Q   W
                                                \label{Ap18}\;, \\
            {\bar Q}^+ &=& W^{-1} Q^+ W
                                                \label{Ap18a}\;, \\
            {\bar P}^+ &=& W^{-1} P^+ {W^{-1}}^T
                                                \label{Ap18b}\;, \\
            {\bar P}'  &=& W^T    P   W
                                                \label{Ap18c}\;,
   \end{eqnarray}\end{mathletters}\noindent%
have to be known.

The use of the canonical basis also facilitates calculation
of transition matrix elements of two-body operators.
Moreover, the particle-number projection discussed in
Sec.{\ }\ref{sec2} can then be performed at a very low cost.
This is so because the multiplication of
the Thouless matrix $C^+$ of
the ket state by the gauge phase factor $e^{2i\phi}$
does not change the canonical
eigenvectors. Only the eigenvalues $D_i$ are multiplied by $e^{2i\phi}$.
Here we only
give expressions for particle-number-projected matrix
elements of separable interactions:
\ifpreprintsty
   \begin{eqnarray}
      \langle C'|\hat Q\hat Q^+\hat P_N|C\rangle
       =& \sum_{ij}\bigg[&{\bar Q}'_{ii}{\bar Q}^+_{jj} D_i D_j R^2_2(i,j)
                                                     \nonumber \\
           &+&            {\bar Q}'_{ij}{\bar Q}^+_{ji} D_j     R^2_1(i,j)
                                                     \nonumber \\
           &+&            {\bar Q}'_{ij}
                          {\bar Q}^+_{\tilde\imath\tilde\jmath}
                                   s_j s_i^{\displaystyle\ast}
                          \sqrt{D_i D_j}R^2_1(i,j)
                   \bigg]
                                                \label{Ap19}\;, \\
      \langle C'|\hat P^+\hat P\hat P_N|C\rangle
       =& \sum_{ij}\bigg[&{\bar P}'_{\tilde\imath i}
                          {\bar P}^+_{j\tilde\jmath}
                        s_i s_j^{\displaystyle\ast}\sqrt{D_i D_j} R^2_1(i,j)
                                                     \nonumber \\
           &+2&           {\bar P}'_{ij}{\bar P}^+_{ji} D_i D_j  R^2_2(i,j)
                                                               \bigg]
                                                \label{Ap20}\;,
   \end{eqnarray}
\else
\widetext
   \begin{eqnarray}
      \langle C'|\hat Q\hat Q^+\hat P_N|C\rangle
       =& \sum_{ij}\bigg[&{\bar Q}'_{ii}{\bar Q}^+_{jj} D_i D_j R^2_2(i,j)
            +             {\bar Q}'_{ij}{\bar Q}^+_{ji} D_j     R^2_1(i,j)
            +             {\bar Q}'_{ij}
                          {\bar Q}^+_{\tilde\imath\tilde\jmath}
                                   s_j s_i^{\displaystyle\ast}
                          \sqrt{D_i D_j}R^2_1(i,j)
                                                               \bigg]
                                                \label{Ap19} , \\
      \langle C'|\hat P^+\hat P\hat P_N|C\rangle
       =& \sum_{ij}\bigg[&{\bar P}'_{\tilde\imath i}
                          {\bar P}^+_{j\tilde\jmath}
                        s_i s_j^{\displaystyle\ast}\sqrt{D_i D_j} R^2_1(i,j)
            +2            {\bar P}'_{ij}{\bar P}^+_{ji} D_i D_j  R^2_2(i,j)
                                                               \bigg]
                                                \label{Ap20} ,
   \end{eqnarray}\narrowtext\noindent\fi%
where $R^2_k(i,j)$ are the residues \cite{DMP64,RS80}
   \begin{equation}\label{Ap21}
   R^2_k(i,j)={1\over 2\pi} \!\int_0^{2\pi}\!\! d\phi\,
              \frac{\langle C'|C(\phi)\rangle e^{-i(N-2k)\phi}}
                   {\left(1+D_ie^{2i\phi}\right)
                    \left(1+D_je^{2i\phi}\right)}\;,
   \end{equation}
which can be calculated by using the discretized integrals
discussed in
Sec.{\ }\ref{sec2}.
The overlap of Thouless states is given by
   \begin{equation}\label{Ap22}
    \langle C'|C(\phi)\rangle
       =\det\left(1+e^{2i\phi}C^+C'\right)^{1/2}
       =\prod_{i>0}\left(1+D_ie^{2i\phi}\right)\;,
   \end{equation}
where notation $i$$>$0 means that only one state from every canonical
pair $(i,\tilde\imath)$ is included in the product.

\section{Matrix elements between time-even states}

Suppose that the Thouless states are time-even, i.e.,
   \begin{equation}\label{Ap23}
     \hat T|C\rangle = |C\rangle\;,
   \end{equation}
where $\hat T$ is the antiunitary time-reversal operator, and suppose
that the single-particle basis is closed with respect to the
time-reversal, i.e.,
   \begin{mathletters}\begin{eqnarray}
    \hat T a^+_m \hat T^+ &=& \sum_{m'} U^T_{mm'}a^+_{m'}\;, \label{Ap24a} \\
    \hat T a_m   \hat T^+ &=& \sum_{m'} U^+_{mm'}a_{m'}  \;, \label{Ap24b}
   \end{eqnarray}\end{mathletters}\noindent%
where $U$ is a unitary antisymmetric matrix. Then the Thouless matrix $C^+$
defining the state $|C\rangle$ is time-even, namely,
   \begin{equation}\label{Ap25}
    C^+ = U C^T U^T \;.
   \end{equation}

For every antisymmetric time-even matrix such as $C^+$ we may define
the hermitian time-even matrix $\tilde C$,
   \begin{equation}\label{Ap26}
    \tilde C = C^+U^{\displaystyle\ast}
             = C^{\displaystyle\ast} U^+ = U C^T = U^T C \;,
   \end{equation}
   \begin{equation}\label{Ap27}
    \tilde C^+=\tilde C\;, \quad\quad \tilde C
              = U \tilde C^{\displaystyle\ast} U^+ \;,
   \end{equation}
which contains the same information about the Thouless state $|C\rangle$
as the matrix $C^+$.
Results of Appendices \ref{AppA} and \ref{AppB} can now be significantly
simplified.

First, the matrix $C^+C'$ defining all transitional matrix
elements becomes quasihermitian,
   \begin{equation}\label{Ap28}
    C^+C'= \tilde C \tilde C'
         = \tilde C^{1/2} \left(\tilde C^{1/2}\tilde C'\tilde C^{1/2}\right)
           \tilde C^{-1/2} \;,
   \end{equation}
i.e., by a similarity transformation $\tilde C^{1/2}$ can be transformed
into a hermitian matrix $\tilde Z$:
   \begin{equation}\label{Ap30}
    \tilde Z = \tilde C^{1/2}\tilde C'\tilde C^{1/2} \;.
   \end{equation}
Therefore, all eigenvalues $D_i$ become real numbers.
(Cases when $\tilde C$ is singular can be considered separately).

Second, the eigenvectors of the hermitian matrix $\tilde Z$
form a unitary matrix $V$,
   \begin{equation}\label{Ap31}
   \tilde Z V = V D \;,
   \end{equation}
and hence the eigenvectors of $C^+C'$
(cf. Eq.{\ }(\ref{Ap1})) are given by
   \begin{equation}\label{Ap29}
    W = \tilde C^{1/2} V \;.
   \end{equation}

Third, the hermitian matrix in Eq.{\ }(\ref{Ap28}) is time-even and
therefore has eigenvalues pairwise degenerate due to the Kramers
degeneracy. Therefore, only half of the eigenvectors have to be
numerically calculated and the canonical pairs of Appendix \ref{AppA}
can be identified with the time-reversed pairs \cite{NW83}.

\begin{table}
\caption[5]{Absolute values of the reduced matrix elements of the quadrupole
            operator between the lowest $2^+$ and $0^+$
            states obtained in the exact
            calculation for $j$=15/2 and $N$=8. }
\label{t5}
\begin{tabular}{clllll}
          &~~$0_{1}^+$&~~$0_{2}^+$&~~$0_{3}^+$&~~$0_{4}^+$&~~$0_{5}^+$\\
\hline
$2_{1}^+$ & 1.921     & 0         & 0.017     &  0.003    &  0        \\
$2_{2}^+$ & 0         & 1.307     & 0         &  0        &  0.014    \\
$2_{3}^+$ & 0         & 1.231     & 0         &  0        &  0.034    \\
$2_{4}^+$ & 0.086     & 0         & 0.790     &  0.208    &  0        \\
$2_{5}^+$ & 0.070     & 0         & 0.387     &  0.437    &  0        \\
\end{tabular}
\end{table}

\begin{table}
\caption[6]{Same as Table \ref{t5} for $j$=31/2 and $N$=4. }
\label{t6}
\begin{tabular}{cr@{}lr@{}lr@{}lr@{}l}
          &   &~~$0_{1}^+$& &~~$0_{2}^+$&   &~~$0_{3}^+$&   &~~$0_{4}^+$\\
\hline
$2_{1}^+$ &   &1.366     & & 0.004    &$<$& 0.001    &$<$& 0.001    \\
$2_{2}^+$ &   &0.099     & & 0.109    &   & 0.001    &$<$& 0.001    \\
$2_{3}^+$ &   &0.005     & & 1.091    &   & 0.018    &   & 0.002    \\
$2_{4}^+$ &   &0.001     & & 0.181    &   & 0.206    &   & 0.015    \\
$2_{5}^+$ &$<$&0.001     & & 0.020    &   & 0.833    &   & 0.072    \\
\end{tabular}
\end{table}

\narrowtext
\begin{table}
\caption[T1]{The decomposition of the $N$=8 SCEM collective states
             (\ref{a1a})
             in the single-$j$ shell for $j$=15/2 into states of
             good angular momentum. The numbers of new states obtained
             for a given value $N_F$ as compared to those for $N_F$$-$1
             are listed for every $I$.}
\label{t1}

\begin{tabular}{c|cccccccccc}
 $I$ &  \multicolumn{10}{c}{$N_F$}\\
   &  \ifpreprintsty
          \multicolumn{10}{c}{\hspace{-35pt}\hrulefill}
      \else
          \multicolumn{10}{c}{\hspace{-14pt}\hrulefill}
      \fi     \\
     & 0 &  1 &  2 &  3 &  4 &  5 &  6 &  7 &  8 &  9 \\
 \hline
   0 & 1 &  0 &  1 &  1 &  1 &  1 &  1 &  1 &  0 &  0 \\
   1 & 0 &  0 &  0 &  0 &  0 &  1 &  2 &  1 &  0 &  0 \\
   2 & 0 &  1 &  1 &  1 &  2 &  3 &  5 &  3 &  0 &  0 \\
   3 & 0 &  0 &  0 &  1 &  1 &  2 &  4 &  4 &  1 &  0 \\
   4 & 0 &  0 &  1 &  1 &  2 &  5 &  7 &  6 &  3 &  0 \\
   5 & 0 &  0 &  0 &  0 &  1 &  2 &  6 &  8 &  4 &  0 \\
   6 & 0 &  0 &  0 &  1 &  1 &  2 &  6 &  3 &  8 &  0 \\
   7 & 0 &  0 &  0 &  0 &  0 &  1 &  2 &  9 & 11 &  3 \\
   8 & 0 &  0 &  0 &  0 &  1 &  1 &  2 &  7 & 15 &  9 \\
   9 & 0 &  0 &  0 &  0 &  0 &  0 &  1 &  3 & 11 & 11 \\
  10 & 0 &  0 &  0 &  0 &  0 &  1 &  1 &  2 &  9 & 15 \\
  11 & 0 &  0 &  0 &  0 &  0 &  0 &  0 &  1 &  3 & 13 \\
  12 & 0 &  0 &  0 &  0 &  0 &  0 &  1 &  1 &  3 & 10 \\
  13 & 0 &  0 &  0 &  0 &  0 &  0 &  0 &  0 &  1 &  4 \\
  14 & 0 &  0 &  0 &  0 &  0 &  0 &  0 &  1 &  1 &  3 \\
  15 & 0 &  0 &  0 &  0 &  0 &  0 &  0 &  0 &  0 &  1 \\
  16 & 0 &  0 &  0 &  0 &  0 &  0 &  0 &  0 &  1 &  1 \\
  17 & 0 &  0 &  0 &  0 &  0 &  0 &  0 &  0 &  0 &  0 \\
  18 & 0 &  0 &  0 &  0 &  0 &  0 &  0 &  0 &  0 &  1 \\
\end{tabular}
\end{table}

\begin{table}
\caption[3]{Maximum quadrupole moments}
\label{t3}
\begin{tabular}{ccc}
   $j$    &  $N$  &  $Q_{\text{max}}$  \\
   \hline
   15/2   &   8   &    1.6      \\
   31/2   &   10  &    1.7      \\
   31/2   &   12  &    2.0      \\
   31/2   &   14  &    2.2      \\
   31/2   &   16  &    2.4      \\
\end{tabular}
\end{table}

\begin{table}
\caption[T2]{Numbers of states $N_{\text{exact}}$ of spin $I$ in the
             single-$j$ shell for $j$=15/2 and $N$=8 particles compared
             with the numbers $k_{\mbox{\scriptsize crit}}$
             of eigenvalues of the overlap matrix retained
             for the 10 and 19 point GCM calculations (see text).}
\label{t2}
\begin{tabular}{cccc}
 $I$  &  $N_{\text{exact}}$ &  $k_{\mbox{\scriptsize crit}}$~(10 points)
                            &  $k_{\mbox{\scriptsize crit}}$~(19 points)  \\
 \hline
 0    &  7   &  7  &   7    \\
 1    &  4   & --  &  --    \\
 2    & 16   & 12  &  16    \\
 3    & 13   &  3  &   7    \\
 4    & 25   & 13  &  21    \\
 5    & 21   &  6  &  12    \\
 6    & 31   & 15  &  22    \\
 7    & 26   &  8  &  14    \\
 8    & 35   & 16  &  24    \\
 9    & 29   & 10  &  16    \\
10    & 35   & 16  &  23    \\
11    & 29   & 11  &  15    \\
12    & 34   & 14  &  18    \\
13    & 27   & 10  &  11    \\
14    & 30   & 10  &  11    \\
15    & 23   &  6  &   6    \\
16    & 25   &  6  &   6    \\
17    & 19   &  3  &   3    \\
18    & 20   &  4  &   4    \\
\end{tabular}
\end{table}

\begin{figure}
\caption[F11]{Exact yrast spectra for $N$=8 and $j$=15/2 obtained
              with varying strength of pairing and quadrupole
              interactions.}
\label{f11}
\end{figure}

\begin{figure}
\caption[F13]{Exact yrast spectra for for different particle
              numbers $N$ and shell angular momenta $j$.}
\label{f13}
\end{figure}

\begin{figure}
\caption[F16]{Single-particle energies in the intrinsic frame for $j$=15/2
              as functions of the quadruploe moment $Q$ (the Nilsson diagram).
              Exchange term in the mean field is neglected.}
\label{f16}
\end{figure}

\begin{figure}
\caption[F2]{Potential energy surfaces (PES) as functions
             of the $\beta$-$\gamma$ collective
        coordinates corresponding to
        $j$=15/2 and $N$=8 (a), and to $j$=31/2 and $N$=10 (b).
        Full dots denote 10 deformation points used in
        discretizing the GCM equation in the intrinsic frame.}
\label{f2}
\end{figure}

\begin{figure}
\caption[F14]{Exact spectrum for $N$=8 particles occupying the $j$=15/2 shell}
\label{f14}
\end{figure}

\begin{figure}
\caption[F10]{Exact spectrum for $N$=4 particles occupying the $j$=31/2 shell}
\label{f10}
\end{figure}

\begin{figure}
\caption[F1]{CPU times required to calculate one hundred
             of overlaps for different Euler angles
             as a function of the shell angular momentum $j$.
             Solid lines show polynomial fits to the PC-486
             and CRAY results,
             0.237$j^3$ + 0.882$j^2$ and
             0.159$j^2$ + 0.314$j$, respectively.
}
\label{f1}
\end{figure}

\begin{figure}
\caption[F15]{Absolute values of eigenvalues of the GCM overlap
              matrix for $j$=15/2 and $N$=8. Results obtained with
              the 19-point mesh are shown
              for $I$=0 (a) and $I$=6 (b).}
\label{f15}
\end{figure}

\begin{figure}
\caption[F3]{GCM spectra of spin $I$=0 states obtained
             for different values of the cut-off parameter $k$.
             Results obtained with 10 (a) and 19 (b)
             mesh points of intrinsic deformations
             are shown for $j$=15/2 and $N$=8.}
\label{f3}
\end{figure}

\begin{figure}
\caption[F4]{Same as Fig.{\ }\ref{f3} for the $I$=2 states.}
\label{f4}
\end{figure}

\begin{figure}
\caption[F7]{GCM spectra of low-spin spin $I$$\leq$6 states
             obtained with 10 (a) and 19 (b)
             mesh points of intrinsic deformations
             shown for $j$=15/2 and $N$=8.}
\label{f7}
\end{figure}

\begin{figure}
\caption[F8]{GCM spectra of low-energy spin $I$$\leq$18 states
             obtained with 10 (a) and 19 (b)
             mesh points of intrinsic deformations
             shown for $j$=15/2 and $N$=8.}
\label{f8}
\end{figure}

\begin{figure}
\caption[F3B]{Same as Fig.{\ }\ref{f3} for $j$=31/2 and $N$=10.}
\label{f3bis}
\end{figure}

\begin{figure}
\caption[F4B]{Same as Fig.{\ }\ref{f4} for $j$=31/2 and $N$=10.}
\label{f4bis}
\end{figure}

\widetext
\begin{figure}
\caption[F9]{GCM spectra for $j$=31/2 obtained with 19
             mesh points of intrinsic deformations for
             particle numbers between $N$=4 and $N$=14, (a) to (f),
             respectively.}
\label{f9}
\end{figure}
\narrowtext


\begin{references}
\bibitem
          {beta} We denote the quadrupole deformation by the traditional
                 symbol $\beta$ but its scale does not correspond to the
                 usual Bohr's variable. In fact, in the single-$j$
                 shell the quadrupole moment (and the deformation)
                 can be scaled arbitrarily because the radial
                 wave function is not specified.
\bibitem
         {spur}   A shown in Ref.{\ }\cite{BDF90},
                  spurious states may
                  appear even if all eigenvalues of the overlap matrix
                  are positive. This can be attributed to the
                  fact that no exact particle
                  number and angular momentum projection has been performed
                  in that study.
\end{references}
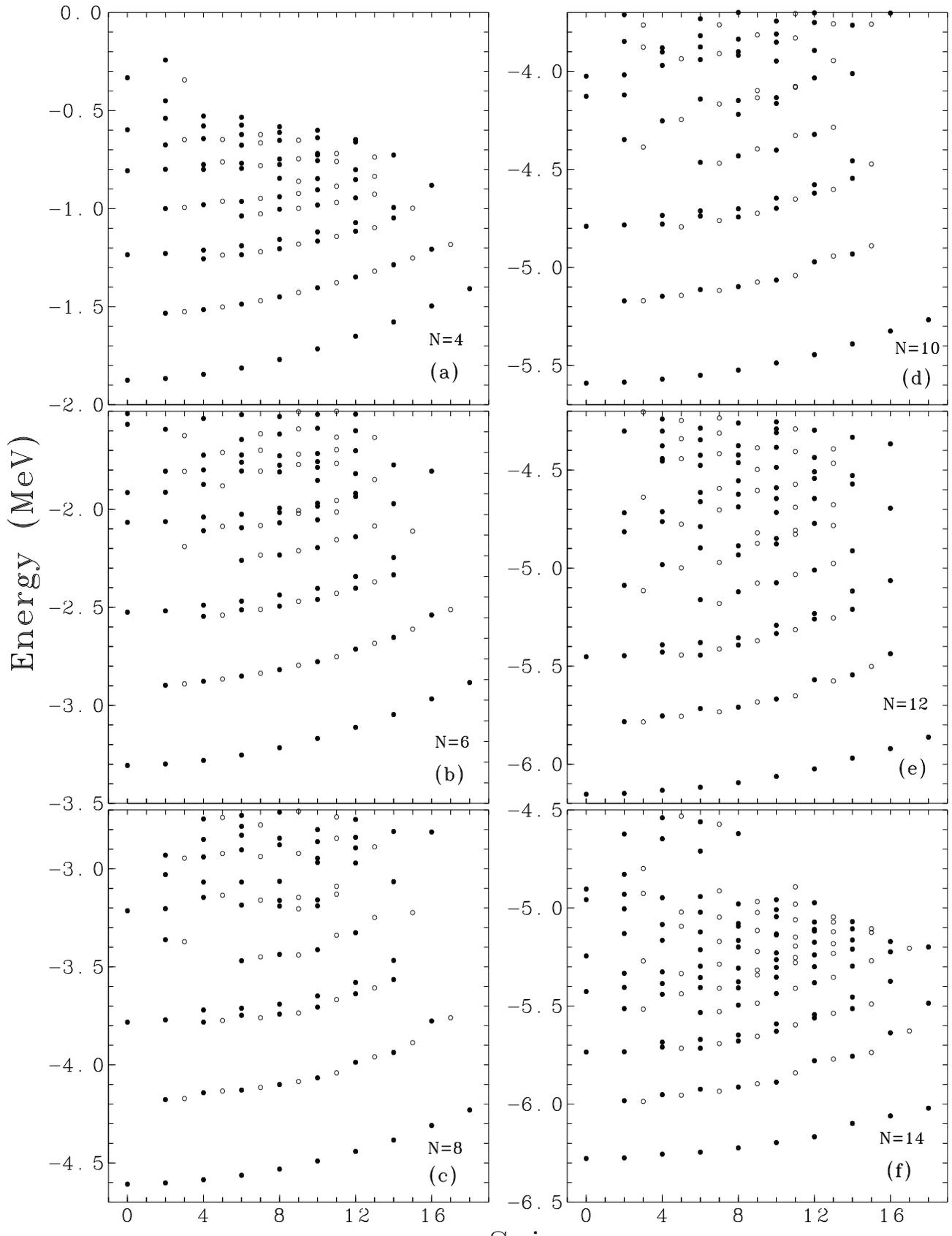
\end{document}